# Unusual Formation of Point Defect Complexes in the Ultra-wide Band Gap Semiconductor $\beta$-Ga$_2$O$_3$

Jared M. Johnson,[1,*] Zhen Chen,[2] Joel B. Varley,[3] Christine M. Jackson,[4] Esmat Farzana,[4] Zeng Zhang,[4] Aaron R. Arehart,[4] Hsien-Lien Huang,[1] Arda Genc,[5] Steven A. Ringel,[4] Chris G. Van de Walle,[6] David A. Muller,[2,7] Jinwoo Hwang[1,*]

[1]Department of Materials Science and Engineering, The Ohio State University, Columbus, Ohio 43210, USA
[2]School of Applied and Engineering Physics, Cornell University, Ithaca, New York 14853, USA
[3]Lawrence Livermore National Laboratory, Livermore, California 94550, USA
[4]Department of Electrical and Computer Engineering, The Ohio State University, Columbus, Ohio 43210, USA
[5]Thermo Fisher Scientific, Hillsboro, Oregon 97124, USA
[6]Materials Department, University of California, Santa Barbara, California 93106, USA
[7]Kavli Institute at Cornell for Nanoscale Science, Cornell University, Ithaca, New York 14853, USA

[*]To whom the correspondence should be addressed.
Electronic mail: johnson.6423@buckeyemail.osu.edu; hwang.458@osu.edu

**Abstract**

Understanding the unique properties of ultra-wide band gap semiconductors requires detailed information about the exact nature of point defects and their role in determining the properties. Here, we report the first direct microscopic observation of an unusual formation of point defect complexes within the atomic scale structure of $\beta$-$Ga_2O_3$ using high resolution scanning transmission electron microscopy (STEM). Each complex involves one cation interstitial atom paired with two cation vacancies. These divacancy – interstitial complexes correlate directly with structures obtained by density functional theory, which predicts them to be compensating acceptors in $\beta$-$Ga_2O_3$. This prediction is confirmed by a comparison between STEM data and deep level optical spectroscopy results, which reveals that these complexes correspond to a deep trap within the band gap, and that the development of the complexes is facilitated by Sn doping through the increase in vacancy concentration. These findings provide new insight on this emerging material's unique response to the incorporation of impurities that can critically influence their properties.

Controlling point defects in crystalline materials can critically influence their properties, and it is therefore imperative to have well-controlled point defects to advance the materials for successful application. Point defects can be very diverse, both in terms of their structure and function. Especially, understanding the formation of point defect complexes, which may occur in response to impurity incorporation within the structure, is of great importance due to their versatile formations and influence on the materials' properties. This is particularly the case in wide-band gap semiconductors, where the intrinsic advantages of a large band gap and the possibility of high optical transparency (e.g. in transparent conductive oxides or TCOs), are severely impaired by the presence of defects. For example, the large critical field strengths that can theoretically be supported in wide-band gap semiconductors for applications in next-generation power electronics can be ruined by the presence of defects [1], and the existence of significant deep-level defect concentrations can severely degrade the optical properties of TCOs [2]. Additionally, the presence of impurities and the formation of various types of complexes in TCOs have been suggested to contribute to the observed intrinsic $n$-type behavior, difficulty in $p$-type doping, and low doping efficiency [3–5]. What has been missing in the field is direct experimental information on the detailed atomic scale structure of such complexes. Gaining this information is essential to exactly crosslinking theoretical predictions of complexes to measured properties of wide band gap semiconductors, which will then provide important guidance to the synthesis and doping of the material with precisely controlled properties. However, such information has been nearly unattainable because of the small (atomic-scale) nature of the defect complexes, especially when they are buried within the 3-dimensional material. These challenges have led to the lack of experimental information on how the complexes incorporate within the atomic scale structure and the inability to discover any unknown complexes that may critically affect the properties of the material.

Here, we present the first direct scanning transmission electron microscopy (STEM) observation of the unusual formation of point defect complexes within the atomic scale structure of $\beta$-Ga$_2$O$_3$. $\beta$-Ga$_2$O$_3$ is an excellent candidate for high performance electronic, optical, power device and sensor applications [6–15] due to its unique properties including an ultra-wide band gap of ~ 4.8 eV [16], optical transparency into the ultraviolet region [17], and high breakdown voltage [18]. However, advancing the material has been hampered by the lack of a detailed understanding of the formation of point defect complexes [4, 19–24] and their impact on electrical and optical properties [25–29]. Using STEM, we discovered a new type of point defect complex that involves one cation interstitial atom, which can be positioned at two of the five possible interstitial sites, paired by two cation vacancies. The structure of this unusual cation interstitial – divacancy complex is consistent with the predictions made by density functional theory (DFT). DFT also shows that this defect acts as a deep level and is the dominant compensating acceptor in $\beta$-Ga$_2$O$_3$; since $\beta$-Ga$_2$O$_3$ is n-type-doped for most applications, this defect therefore has a crucial impact on device performance. Comparing the STEM data to deep level optical spectroscopy (DLOS) shows that formation of the defects is enhanced by increased Sn doping, which increases the Fermi level energy and subsequently increases the vacancy concentration due to formation energy arguments, confirming the compensating-acceptor character, with a defect level at $E_C$ – 2.0 eV. The present atomic scale investigation identifies this unusual defect as the origin of a number of previously unexplained phenomena in $\beta$-Ga$_2$O$_3$, and also sheds light on this material's unique response to the incorporation of impurities that can critically influence its properties.

First, we explain the observation of interstitial defects and defect complexes in Sn-doped $\beta$-Ga$_2$O$_3$ bulk crystals (carrier concentration of 8.2 x 10$^{18}$ cm$^{-3}$ [30]). The unit cell of monoclinic $\beta$-Ga$_2$O$_3$ contains two crystallographically different Ga (Ga$_1$, Ga$_2$), and three oxygen atom positions (O$_1$, O$_2$, O$_3$) as shown in Fig. 1(a). The structure is oriented along [010]$_m$, which is the orientation used throughout this study. Ga$_1$ and Ga$_2$ have tetrahedral and octahedral coordination, respectively. O$_1$ and O$_2$ have threefold coordination, while O$_3$ has fourfold coordination. Figure 1(a) also shows five potential cation interstitial sites (i$_{a-e}$). As will be explained in detail, these interstitial sites have been derived from both our experimental observation and DFT calculations [19, 21]. A high angle annular dark field (HAADF) STEM image (See Methods for details) from the [010]$_m$ Sn-doped sample of a defect free region is shown in Fig. 1(b). As HAADF intensity depends on the atomic number, high intensity in this image mostly arises from the scattering of the Ga$_1$ and Ga$_2$ columns while only very weak intensity is observed from O positions. Figure 2 shows the direct detection of interstitial defects and defect complexes in the same Sn-doped $\beta$-Ga$_2$O$_3$ bulk crystal. In the sample areas shown in Fig. 2(a) (left and right), significant intensities were observed in multiple interstitial sites, in addition to Ga columns that are still positioned at their regular positions. Specifically, these noticeable intensities appeared in four distinct interstitial sites, termed i$_{b-e}$ [Figs. 2(b)-2(e)]. Amongst them, the most prominent intensities were located in the i$_c$ site [Figs. 2(c) and 2(f)]. We also note that these interstitials were found to be clustered and concentrated in ~ 10 nm$^2$ areas [Fig. 2(g)]. Clustering likely happens along the direction parallel to the electron beam as well, which can be evidenced by the high interstitial column intensity (e.g. Fig. 2(f)) indicating that there are likely multiple cation interstitial atoms along the column [31, 32]. Previous works have identified extended defects (*e.g.* twin boundaries and screw dislocations) [33–35] and some atomic scale defects [2, 36] in *β*-Ga$_2$O$_3$, but our present data from aberration corrected STEM provides the first direct identification of the exact positions of interstitial defects.

Next, we identify the unique formation of point defect complexes from the HAADF-STEM intensities by directly correlating them with DFT calculations (See Methods for details). DFT calculations [19, 20, 24] have shown that cation vacancies have low formation energies under O-rich growth conditions, with tetrahedral $V_{Ga}^1$ being the most favorable. However, the vacancy on the tetrahedral Ga$_1$ site was identified as metastable; the presence of the vacancy causes a neighboring Ga atom to leave its tetrahedral site (creating a second vacancy) and move towards an interstitial site midway between the two vacancies, effectively resulting in a $2V_{Ga}^1$ – Ga$_i$ complexes [Fig. 3]. This complex is lower in energy than the isolated vacancy, and acts as a deep acceptor. Figure 3(a) shows how the adjacent Ga$_1$ atom relaxes to the i$_c$ site, becoming octahedrally coordinated and creating an additional $V_{Ga}^1$, which can be formed easily due to the relatively small energy barrier [19, 21]. These DFT results have also been utilized in recent *β*-Ga$_2$O$_3$ defect studies using electron spin resonance [37, 38] and infrared spectroscopy [39, 40] to attribute the presence of proton irradiation induced defects and implanted O-H bonds to the same point defect complexes. However, the direct confirmation of $2V_{Ga}^1$ – $Ga_i^c$ complexes by STEM in *β*-Ga$_2$O$_3$ requires the detection of not only the cation interstitial atoms (explained above) but also the vacancies in the neighboring Ga columns. Identifying a single vacancy within an atomic column may not be trivial [41], however, the presence of several vacancies within a Ga column will decrease the overall HAADF STEM signal due to the loss of scattering from the absent Ga atoms [42]. Image simulations performed using the multislice method [43] (see Methods for details) (See Methods for details) confirmed that signal increases at interstitial sites from cation interstitials and decreases at Ga$_1$ sites from vacancies [Fig. 3(b-bottom) to 3(d-bottom)]. The loss of intensity for Ga$_1$ columns neighboring interstitials was in fact observed in our experimental data. Individual line profiles in Fig. 3(d-top) illustrate the reduced intensity of $Ga_1$ columns adjacent to interstitials [Fig. 3(c-top)] compared to $Ga_1$ columns in a defect free region [Fig. 3(b-top)]. The reduced intensity of $Ga_1$ columns adjacent to interstitials was substantially more significant than the overall intensity decrease in the area (including the $Ga_2$ columns that are not expected to be involved in the relaxation process) that may be caused by the strain [44] due to the interstitials (see Supplemental Material, Table S1). Image simulations indicate that the effect of electron channeling from interstitials is minimal due to the considerable gap (~0.18 nm) between the interstitials and the adjacent Ga columns. Therefore, the substantial intensity decrease apart from the overall decrease due to strain in the $Ga_1$ columns adjacent to interstitials suggests that vacancies are most likely present in those columns. Similarly, vacancy – interstitial complexes surrounding $i_b$ sites ($2V_{Ga}^1 - Ga_i^b$) were also observed [Figs. 3(e)-3(h)]. Quantitative comparison of atomic column intensity to the image simulation can provide the information on the number of defects located within each column (*e.g.* Ref. 31), is highly dependent on the number, location and chemical species of the defect complexes (see Supplemental Material, Fig. S1).

Considering the near-equilibrium growth condition of the edge-defined, film-fed growth (EFG) process used to grow these bulk crystals, the formation energy ($i_c < i_b < i_a$), predicted by DFT [19], may directly correspond to the number density of each of those point defects within the material. Albeit small sampling in STEM images (such as in Fig. 2(g)) the number densities of interstitials observed in the sampled regions ($i_c > i_b > i_a$) appear to be consistent with the energetic favorability found in the DFT calculation. In fact, we have not observed any interstitial $i_a$ intensity above the threshold value in the sampled areas, which suggests the formation of $i_a$ interstitials may be much less likely than $i_b$ or $i_c$.

Although lesser in signal and fewer in total number, the interstitials observed in sites $i_d$ and $i_e$ found in Sn-doped $β$-Ga$_2$O$_3$ display unexplored interstitial complexes that may be important in understanding the material's properties. Interestingly, interstitial intensities in $i_d$ and $i_e$ sites were always observed to be accompanied by interstitial intensities in nearby $i_b$ and $i_c$ sites [Figs. 2(d) and (e)]. This may imply that $2V_{Ga}^1 - Ga_i^{b,c}$ complexes may facilitate the formation of other complexes that occupy the $i_d$ and $i_e$ sites. The exact formation mechanism of $i_d$ and $i_e$ defects and their implication to the properties remain to be further explored in the near future.

The next question is then how Sn doping affects the formation of the interstitial point defect complexes shown above. In general, Sn impurities have been shown to contribute to the *n*-type behavior of $β$-Ga$_2$O$_3$, producing controllable carrier concentrations from $10^{16}$ up to $10^{19}$ cm$^{-3}$ [30, 45]. Formation energy calculations have indicated that Sn can easily incorporate into $β$-Ga$_2$O$_3$ acting as a shallow donor and preferring to substitute on the octahedrally coordinated Ga$_2$ site [20, 46]. In fact, our STEM investigation has revealed several of these Sn substitutional atoms based on the high atomic column intensity in HAADF mode (*e.g.* Fig. 4(a)). In addition to incorporating as substitutional dopants, Sn, when in high concentration, may promote the formation of the vacancy – interstitial complexes shown above. To verify this hypothesis, we first investigated unintentionally doped (UID) $β$-Ga$_2$O$_3$ (carrier concentration of 2.4 x $10^{17}$ cm$^{-3}$ [30]) using HAADF STEM. UID $β$-Ga$_2$O$_3$ images revealed cation interstitials in the $i_c$ site [Fig. 4(b)], which implies that the interstitials may also be created via the same $V_{Ga}^1$ migration mechanism explained above. A larger sampling resulted in the conclusion that these complexes exist lesser in quantity and smaller in intensity in comparison to Sn-doped $β$-Ga$_2$O$_3$. This is consistent with the expectations that the incorporation of Sn donors that drive the material more *n*-type simultaneously stimulates the formation of the compensating acceptor species like the $2V_{Ga}^1 - Ga_i$ complexes [19, 21, 24] through the subsequent increase in vacancy concentration. Additionally, Sn may also incorporate as the cation interstitial, forming $2V_{Ga}^1$ – $Sn_i$ complexes. Its exact role will be discussed among the succeeding results.

DLOS can reveal the rich spectrum of defect states throughout the large band gap [29], by using monochromatic incident light as a function of energy to photo-emit trapped carriers to a band edge, which enables the determination of energy levels, concentrations, and optical cross-sections of deep states (see Methods for details [29, 47, 48]). Accompanying our STEM results and DFT calculations, the DLOS results are consistent with the role of Sn dopant concentration and the number of vacancy – interstitial complexes present. Most importantly, the concentration of the trap detected at $E_C$ – 2.0 eV shows a positive correlation with Sn concentration [Fig. 4(c)] and the trap has been shown to act as a deep acceptor [49]. This defect behavior is consistent with the electronic state of the $2V_{Ga}^1$ – $Ga_i$ complexes predicted by DFT, where levels associated with the expected $\varepsilon(-2/-3)$ charge-state transition levels have been predicted to fall between $E_C$ – 1.9 and $E_C$ – 2.5 eV [19, 24]. The analogous $\varepsilon(-/-2)$ levels calculated for the $2V_{Ga}^1$ – $Sn_i$ complexes are found to be in a similar energy range, at $E_C$ – 2.34 and $E_C$ – 2.89 eV for the $i_b$ and $i_c$ sites, respectively. The apparent defect concentrations obtained from DLOS represent the average over large areas (as opposed to the relatively smaller area observations of STEM) and are limited by the finite time response of the defects [49]. Nonetheless, the positive correlation between the DLOS detected trap and doping concentration, along with the detection of this trap in UID $\beta$-Ga$_2$O$_3$, directly coincide with the observation of interstitials by STEM imaging, which affirms the physical source of the $E_C$ – 2.0 eV defect state to be the divacancy – interstitial complex.

It would be highly valuable to be able to distinguish whether the observed entities are $2V_{Ga}^1$ – $Ga_i$ (spatially separated from donor impurities), or actual $2V_{Ga}^1$ – $Sn_i$ complexes, since it could help identify strategies for controlling the compensating acceptor. The proposed $2V_{Ga}^1 - Sn_i$ acceptor complexes have also been suggested to contribute to compensation in Sn-doped $β$-Ga$_2$O$_3$ [50]. Our calculations show these complexes can readily form when V$_{Ga}$ are nearby (calculated barriers < 0.3 eV for a $Sn_{Ga}^1$ to displace to octahedrally-coordinated interstitials adjacent to V$_{Ga1}$) and are quite stable, with calculated binding energies of at least 1.4 eV relative to isolated V$_{Ga}$ and Sn$_{Ga}$ species as referenced $2V_{Ga}^1 - Sn_i$ and $Sn_{Ga}^2$ [Fig. 4(d)]. However, we would expect that these complexes would need to form during growth, as Sn is not expected to appreciably incorporate on the tetrahedral sites that would be a prerequisite for their formation post-growth [20]. As mentioned above, determining the species of cation interstitials in Sn-doped $β$-Ga$_2$O$_3$ via HAADF-STEM is difficult because variations in number and location make Ga and Sn interstitials indistinguishable (see Supplemental Material, Fig. S1). However, because DLOS results reveal the same trap (E$_C$ – 2.0 eV) in both UID and Sn-doped $β$-Ga$_2$O$_3$ and DFT calculations predict different electronic states for $2V_{Ga}^1 - Ga_i$ and $2V_{Ga}^1 - Sn_i$ complexes, the observed compensating acceptors correlate with native $2V_{Ga}^1 - Ga_i$ complexes.

In summary, the STEM results provide the first observation of unusual divacancy – cation interstitial complexes, and the structure of the identified complexes matches with DFT predictions that classify them as compensating acceptors. DLOS results further validated the formation of these complexes as the detected E$_C$ – 2.0 eV defect state showed a positive correlation with Sn doping. Sn was determined to facilitate the divacancy – interstitial complex development and possibly incorporate as the cation interstitial. These STEM results provide new important insight on the material's unique response to the impurity incorporation that can significantly affect their properties, which can ultimately offer important guidance to the development of growth and doping of TCOs for novel applications. The observed complex is in essence a low-symmetry configuration of a cation vacancy. It will be highly interesting to explore, using computational theory and/or microscopy, whether vacancies in other materials can also spontaneously undergo such symmetry-breaking distortions.

**Acknowledgement**

S. A. R., C. V., and J. H. acknowledge support by the Department of Defense, Air Force Office of Scientific Research GAME MURI Program (Grant No. FA9550-18-1-0479). This work was performed in part at the Cornell PARADIM Electron Microcopy Facility, as part of the Materials for Innovation Platform Program, which is supported by the NSF under DMR-1539918 with additional infrastructure support from DMR1719875 and DMR-1429155. S. A. R. acknowledges additional support from HDTRA1-17-1-0034. DFT work was partially performed under the auspices of the U.S. DOE by Lawrence Livermore National Laboratory under contract DE-AC52-07NA27344, and partially supported by the Critical Materials Institute, an Energy Innovation Hub funded by the U.S. DOE, Office of Energy Efficiency and Renewable Energy, Advanced Manufacturing Office.

**Appendix: Methods**

($\bar{2}01$) Sn-doped and unintentionally doped (UID) $\beta$-Ga$_2$O$_3$ bulk crystals used in this investigation were fabricated by the edge-defined, film-fed growth (EFG) process by the Tamura Corp. The Sn-doped and UID samples have carrier concentrations of 8.2 x 10$^{18}$ cm$^{-3}$ and 2.4 x 10$^{17}$ cm$^{-3}$, respectively. Crystal orientations such as the [010]$_m$ and [001]$_m$ yield advantageous viewing directions for point defect imaging due to the large atomic spacings. Cross-sectional [010]$_m$ TEM samples were prepared using focused ion beam (FIB). As atomic resolution imaging of $[010]_m$ $\beta$-$Ga_2O_3$ requires thin, clean samples, we further milled TEM samples using a low-energy (500 eV) ion mill (Fischione Nanomill). The final thickness of the TEM foils was determined by position averaged convergent beam electron diffraction to be ~ 25 nm [31, 54], meaning each gallium column contains ~ 80 atoms along the depth direction. STEM was then performed using aberration-corrected Thermo Fisher Scientific Titan Themis STEM instruments, all operated at 300 kV.

All STEM HAADF image simulations were performed using the multislice algorithm [43]. For each image simulation of randomly distributed defect complexes, ten random arrangements were simulated and then averaged to increase statistical reliability. Thermal vibrations were ignored as their contribution to column intensity is minimal. The image simulations used aberration parameters of our probe corrected FEI Titan STEM ($C_{s3}$=0.002 mm, $C_{s5}$=1.0 mm) with a 20.0 mrad convergence half angle at an acceleration voltage of 300 kV.

The DFT atomistic simulations of the vacancy complex energetics were based on hybrid functional calculations performed using the same methodology as described in Ref [24]. Specifically, the vacancy and vacancy complexes were modelled within a 160-atom supercell representation of bulk β-Ga2O3, where corrections to the formation energies of charged defects owing to image-charge interactions were included as detailed in Ref [24]. Migration barriers were computed between linearly interpolated structures and thus offer upper bounds to the real barriers.

DLOS was performed on Ni/$Ga_2O_3$ Schottky diodes on four different β-$Ga_2O_3$ substrates from Tamura: n~1 x $10^{17}$ cm$^{-3}$ UID (010), n~1.5 x $10^{18}$ cm$^{-3}$ Sn-doped (-201), n~3.5 x $10^{18}$ cm$^{-3}$ Sn-doped (010), n~5 x $10^{18}$ cm$^{-3}$ Sn-doped (010). Carrier concentrations on these samples were confirmed by C-V measurements. DLOS utilizes monochromatic sub-bandgap light to observe optically stimulated photoemission transients as a function of incident light energy. Here, light from a 1000W Xe-lamp was dispersed through a high-resolution monochromator to provide incident light from 1.2 eV to 5.0 eV in 0.02 eV steps, and photoemission transients were measured for 300 seconds.

**Figure captions**

Fig. 1. Crystal structure and experimental image of $\beta$-Ga$_2$O$_3$. (a) Crystal structure of $\beta$-Ga$_2$O$_3$ along the [010]$_m$ direction with five possible interstitial sites (i$_{a-e}$). (b) Atomic resolution HAADF-STEM image of a [010]$_m$ Sn-doped $\beta$-Ga$_2$O$_3$ bulk crystal from a defect free area.

Fig. 2. Direct detection of interstitial defects in Sn-doped $\beta$-Ga$_2$O$_3$. (a) HAADF-STEM images from two regions (left and right) with clustered interstitial defects. Magnified locations from image (a-left) with intensity located in interstitial sites (b) i$_b$, (c) i$_c$, (d) i$_d$ and (e) i$_e$ marked corresponding to Fig. 1(a). Arrows in (d) and (e) point towards accompanying interstitials in the i$_b$ and i$_c$ sites. (f) Intensity is evaluated across the red-dashed line demonstrating the significant signal scattered from the cation interstitial atoms located in the i$_c$ site (red triangle). (g) Positions of i$_b$ (aqua mark) and i$_c$ (red mark) interstitial atoms identified in a larger area of $\beta$-Ga$_2$O$_3$. Yellow-dashed outlined areas indicate the interstitials clustering and concentrating in ~ 10 nm$^2$ areas.

Fig. 3. Migration mechanism of $V_{Ga}^1$ for the formation of $2V_{Ga}^1$ - Ga$_i$ complexes. In the presence of a $V_{Ga}^1$, an adjacent Ga$_1$ relaxes into the (a) i$_c$ or the (e) i$_b$ site, creating an additional $V_{Ga}^1$. (b-d) Experimental and simulated HAADF images showing multiple $2V_{Ga}^1$ - Ga$_i^c$ complexes along the [010]$_m$ direction. (d-top) Line profiles from a (b-top, green-dashed) defect free and (c-top, red-dashed) interstitial containing experimental images show a significant reduction in Ga$_1$ intensity (blue arrows) from vacancies when neighboring i$_c$ interstitials (red triangle) compared to unperturbed Ga$_1$ columns (green arrows). (d-bottom) Line profiles from a (b-bottom, green-dashed) simulated defect free and a (c-bottom, red-dashed) simulated 25 nm (~82 atoms) thick crystal containing multiple defect complexes showing similar profiles to the corresponding experimental images. (f-h) Similarly, experimental and simulated HAADF images show multiple $2V_{Ga}^1$ - Ga$_i^b$ complexes along the depth direction.

Fig. 4. STEM, DFT, and DLOS results for point defects and defect complexes in $\beta$-Ga$_2$O$_3$. (a) HAADF STEM image of an atomic column containing Sn dopants (purple triangle) in [010]$_m$ bulk Sn-doped $\beta$-Ga$_2$O$_3$. The inset shows the line profile along the dashed line in the figure. (b) HAADF STEM image of UID $\beta$-Ga$_2$O$_3$ showing interstitial defects with red arrows indicating detected Ga interstitials in the i$_c$ sites. (c) DLOS as a function of Sn doping in bulk $\beta$-Ga$_2$O$_3$. The trap detected at E$_C$ – 2.0 eV shows a positive correlation with Sn concentration. (d) Formation energies for $Sn_{Ga}^2$ and $2V_{Ga}^1 - Sn_i$ complexes vacancies in $\beta$-Ga$_2$O$_3$, shown in the limit of (a) O-rich and (b) Ga-rich conditions.

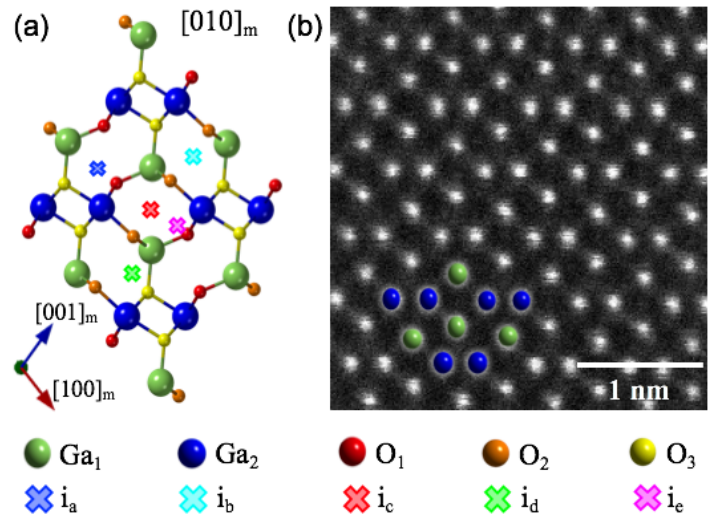

Fig. 1. Crystal structure and experimental image of $\beta$-Ga$_2$O$_3$. (a) Crystal structure of $\beta$-Ga$_2$O$_3$ along the $[010]_m$ direction with five possible interstitial sites ($i_{a-e}$). (b) Atomic resolution HAADF-STEM image of a $[010]_m$ Sn-doped $\beta$-Ga$_2$O$_3$ bulk crystal from a defect free area.

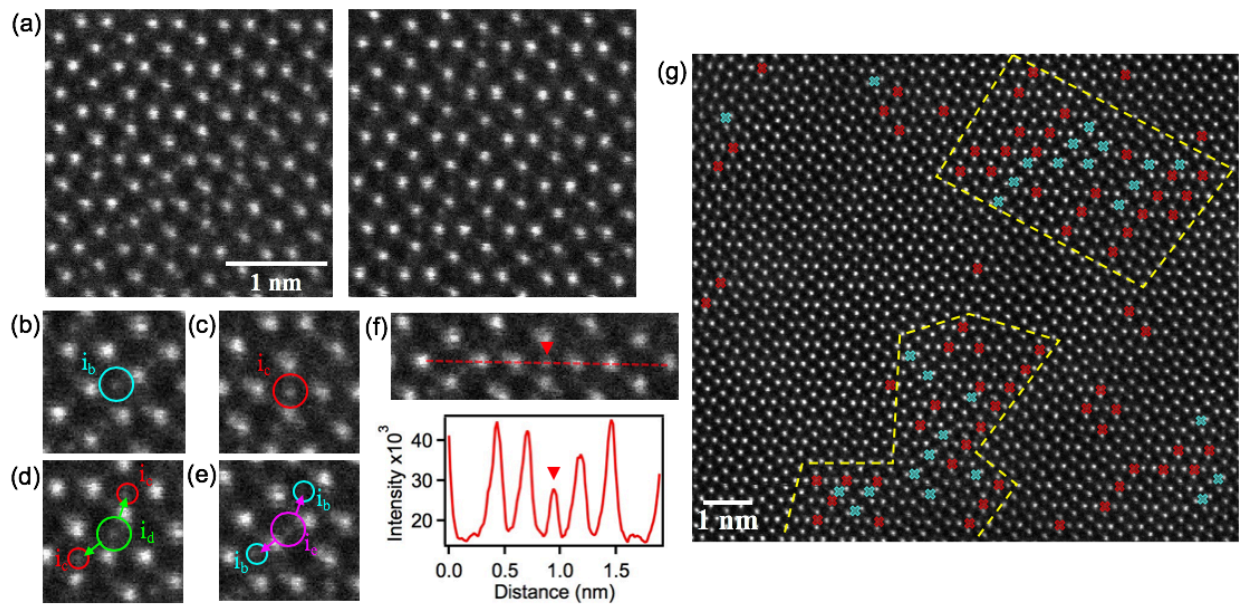

Fig. 2. Direct detection of interstitial defects in Sn-doped $\beta$-Ga$_2$O$_3$. (a) HAADF-STEM images from two regions (left and right) with clustered interstitial defects. Magnified locations from image (a-left) with intensity located in interstitial sites (b) $i_b$, (c) $i_c$, (d) $i_d$ and (e) $i_e$ marked corresponding to Fig. 1(a). Arrows in (d) and (e) point towards accompanying interstitials in the $i_b$ and $i_c$ sites. (f) Intensity is evaluated across the red-dashed line demonstrating the significant signal scattered from the cation interstitial atoms located in the $i_c$ site (red triangle). (g) Positions of $i_b$ (aqua mark) and $i_c$ (red mark) interstitial atoms identified in a larger area of $\beta$-Ga$_2$O$_3$. Yellow-dashed outlined areas indicate the interstitials clustering and concentrating in ~ 10 nm$^2$ areas.

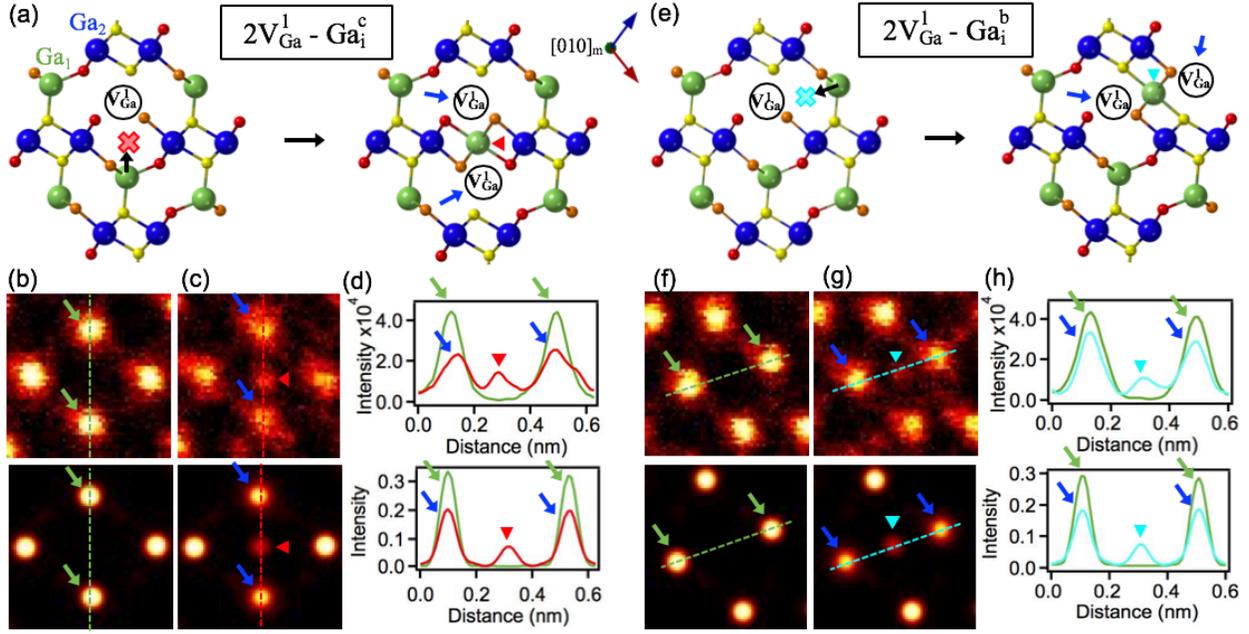

Fig. 3. Migration mechanism of $V_{Ga}^1$ for the formation of $2V_{Ga}^1$ - $Ga_i$ complexes. In the presence of a $V_{Ga}^1$, an adjacent $Ga_1$ relaxes into the (a) $i_c$ or the (e) $i_b$ site, creating an additional $V_{Ga}^1$. (b-d) Experimental and simulated HAADF images showing multiple $2V_{Ga}^1$ - $Ga_i^c$ complexes along the $[010]_m$ direction. (d-top) Line profiles from a (b-top, green-dashed) defect free and (c-top, red-dashed) interstitial containing experimental images show a significant reduction in $Ga_1$ intensity (blue arrows) from vacancies when neighboring $i_c$ interstitials (red triangle) compared to unperturbed $Ga_1$ columns (green arrows). (d-bottom) Line profiles from a (b-bottom, green-dashed) simulated defect free and a (c-bottom, red-dashed) simulated 25 nm (~82 atoms) thick crystal containing multiple defect complexes showing similar profiles to the corresponding experimental images. (f-h) Similarly, experimental and simulated HAADF images show multiple $2V_{Ga}^1$ - $Ga_i^b$ complexes along the depth direction.

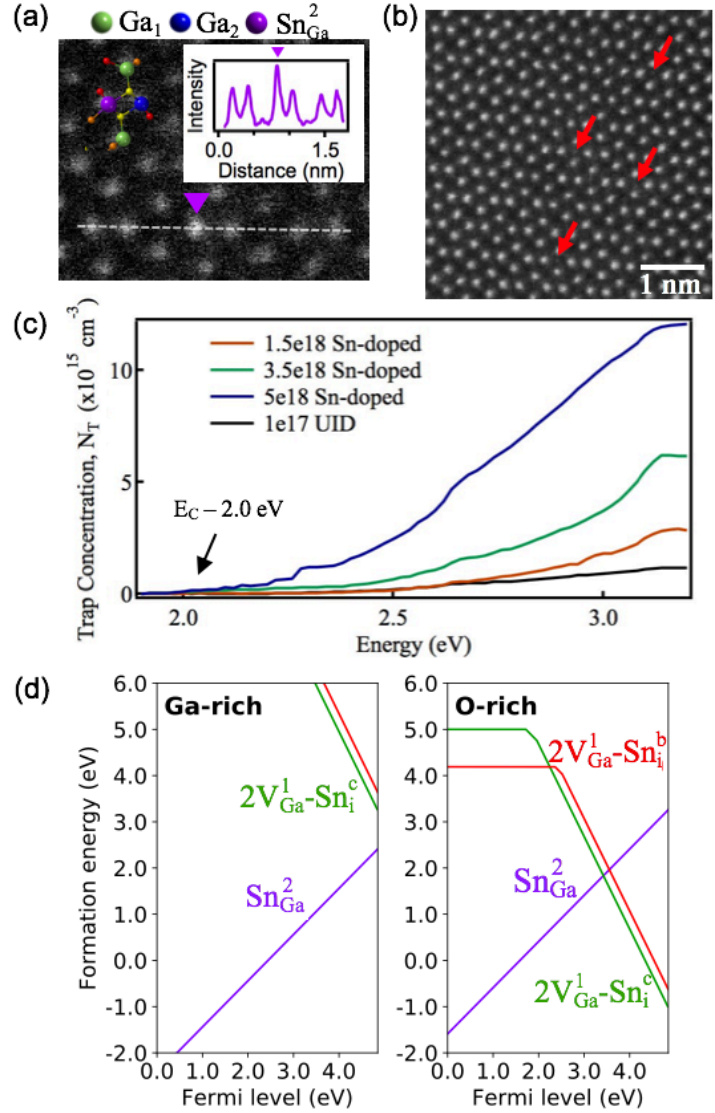

Fig. 4. STEM, DFT, and DLOS results for point defects and defect complexes in $\beta$-Ga$_2$O$_3$. (a) HAADF STEM image of an atomic column containing Sn dopants (purple triangle) in [010]$_m$ bulk Sn-doped $\beta$-Ga$_2$O$_3$. The inset shows the line profile along the dashed line in the figure. (b) HAADF STEM image of UID $\beta$-Ga$_2$O$_3$ showing interstitial defects with red arrows indicating detected Ga interstitials in the i$_c$ sites. (c) DLOS as a function of Sn doping in bulk $\beta$-Ga$_2$O$_3$. The trap detected at $E_C - 2.0$ eV shows a positive correlation with Sn concentration. (d) Formation energies for $Sn_{Ga}^2$ and $2V_{Ga}^1 - Sn_i$ complexes vacancies in $\beta$-Ga$_2$O$_3$, shown in the limit of (a) O-rich and (b) Ga-rich conditions.

# Supplemental Materials

**Unusual Formation of Point Defect Complexes in the Ultra-wide Band Gap Semiconductor $\beta$-Ga$_2$O$_3$**

Jared M. Johnson,[1,*] Zhen Chen,[2] Joel B. Varley,[3] Christine M. Jackson,[4] Esmat Farzana,[4] Zeng Zhang,[4] Aaron R. Arehart,[4] Hsien-Lien Huang,[1] Arda Genc,[5] Steven A. Ringel,[4] Chris G. Van de Walle,[6] David A. Muller,[2,7] Jinwoo Hwang[1,*]

| Neighboring Ga column Type | Avg. Intensity |
|---|---|
| Ga$_1$ when there is no interstitial | 30,164 ± 4.7% |
| Ga$_2$ when there is no interstitial | 31,051 ± 9.2% |
| Ga$_1$ with an interstitial in the i$_b$ site | 23,577 ± 11.5% |
| Ga$_2$ with an interstitial in the i$_b$ site | 26,953 ± 14.6% |
| Ga$_1$ with an interstitial in the i$_c$ site | 22,782 ± 12.5% |
| Ga$_2$ with an interstitial in the i$_c$ site | 25,893 ± 14.0% |

Table S1. Ga$_1$ and Ga$_2$ columns from a defect free region are compared to neighboring Ga$_1$ and Ga$_2$ columns when an interstitial is located in the i$_b$ or i$_c$ site. Average column intensities are calculated by averaging the intensities located within a circle of radius = 8 pixels centered around the column. The neighboring Ga columns with interstitials present were chosen for this analysis from the indicated interstitials in Fig. 2(g).

Confirming the existence of vacancies in Ga$_1$ sites requires identifying the intensity decrease of Ga$_1$ columns in the presence of interstitials while Ga$_2$ columns remain unaffected. Table 1 shows the reduced intensity of Ga$_1$ columns adjacent to interstitials was substantially more significant than the overall intensity decrease in the area (Ga$_1$ vs Ga$_2$ with interstitials). As Ga$_2$ columns are not expected to be involved in the formation of $2V_{Ga}^1$ - $Ga_i$ complexes, this areal decrease in intensity is likely due to strain near interstitials. Despite this loss in intensity, the decrease in $Ga_1$ intensity indicates the presence of vacancies within the columns and the formation of $2V_{Ga}^1$ - $Ga_i$ complexes.

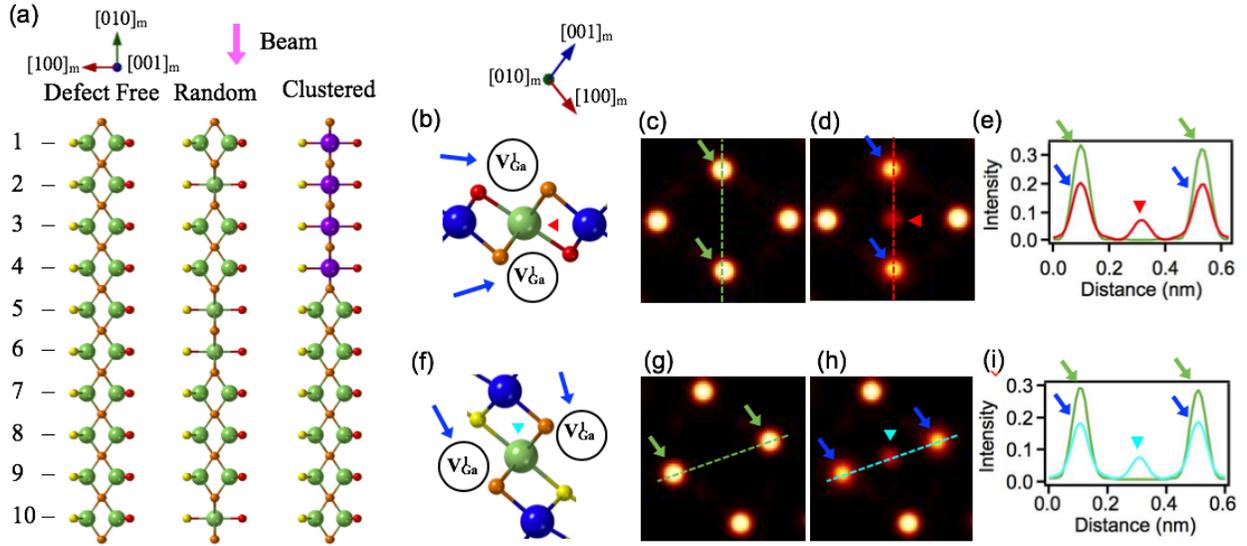

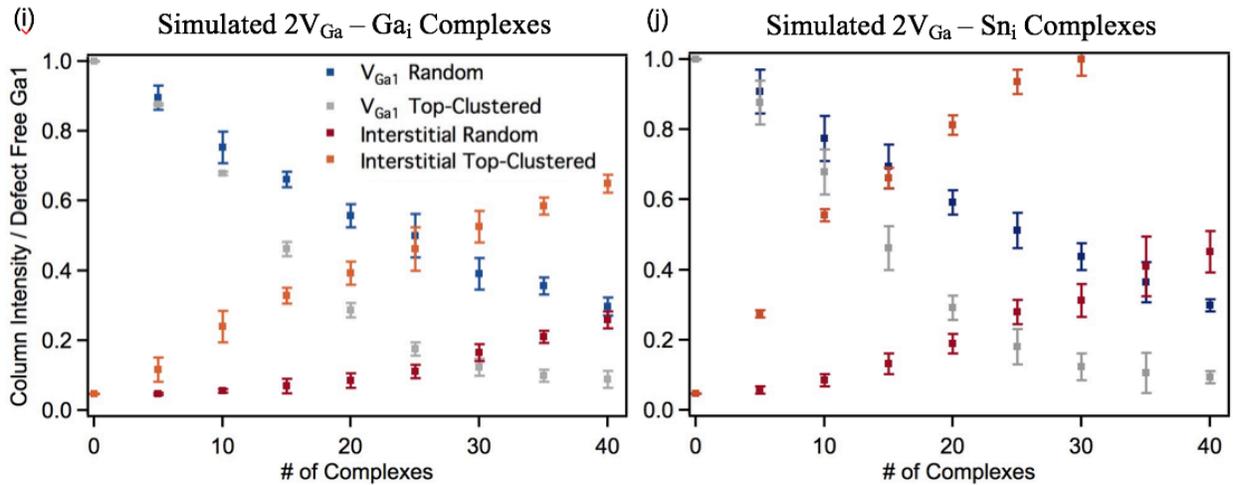

Fig. S1. HAADF multislice image simulations and analysis for $2V_{Ga}^1 - Ga_i$ and $2V_{Ga}^1 - Sn_i$ complexes. (a) Schematic examples of $\beta$-Ga$_2$O$_3$ used in the simulations along the $[010]_m$ beam direction of (a-left) a defect free cell, (a-middle) a cell containing four randomly distributed $2V_{Ga}^1 - Ga_i$ complexes (positions 2,5,6,10), and (a-right) a cell containing four clustered $2V_{Ga}^1 - Sn_i$ complexes (positions 1,2,3,4). (b) Schematic of the $2V_{Ga}^1 - Ga_i^c$ defect complex. (e) Line profiles from a defect free (c, green-dashed) and a $2V_{Ga}^1 - Ga_i^c$ defect complex containing (d, red-dashed) simulated HAADF image showing the decrease in Ga$_1$ intensity (blue arrows) from the incorporated vacancies and increase in interstitial site intensity (red triangle). (f) Schematic of the $2V_{Ga}^1 - Ga_i^b$ defect complex. (i) Similarly, line profiles from a defect free (g, green-dashed) and a $2V_{Ga}^1 - Ga_i^b$ defect complex containing (h, aqua-dashed) simulated HAADF image show similar intensity changes. Plots (i) and (j) show the multislice image simulation analysis for (i) $2V_{Ga}^1 - Ga_i$ and (j) $2V_{Ga}^1 - Sn_i$ complexes in a 25 nm (82 atom) thick crystal. Displayed is the fraction of intensity with respect to a defect free Ga$_1$ column versus the number of complexes inserted along the depth direction. Each plot shows a random arrangement of defect complexes and defect complexes clustered at the surface of the crystal.

Determining the total number of defect complexes located along the depth direction of the TEM sample and identifying the chemical species of the interstitial atoms must involve quantitatively relating the change in intensities to the exact number and location of defect complexes through image simulation. Using the multislice method [43], we have simulated arrangements of defect complexes with both Ga and Sn interstitials (Fig. S1). Trends found in Fig. S1(i) and S1(j) confirm that intensity increases due to interstitials and decreases due to vacancies. By simulating top-surface-clustered and randomly distributed defect complexes, we have obtained the upper and lower limits of possible intensity variation from the defect complexes (due to probe channeling and oscillation, Ref. 31). The analysis indicates that there is a wide range of possible intensity changes due to the immense dependence on location and chemical species of the defect complexes. Thus, determining the exact number and interstitial species of the defect complexes can be problematic. However, despite the challenges mentioned above, insights have been gained from the data. Random distributions of defect complexes along the column display minuscule changes in interstitial intensity (~ 2% for 10 Ga interstitial atoms), which is substantially smaller than the intensity change that observed in the experimental data. This means that a random distribution would require the defect segregation involve many atoms, extending beyond a few nanometers, which would be inconsistent to observed clustering along the lateral direction shown in Fig. 2(g). This weighs more on the scenario that the cations are in fact clustered within the column, which would not be surprising, given that the clustering is already observed along the lateral direction. Assuming the top-surface-clustering, the number of cation interstitials is somewhere in between 5 and 10 depending on the species being Ga or Sn.